\documentclass{aa}  
\usepackage{graphicx}
\usepackage{natbib}
\bibpunct{(}{)}{;}{a}{}{,}
\begin{document}

\title{Cometary water expansion velocity from OH line shapes}

\titlerunning{Cometary water expansion velocity}      
   \author{W.-L.~Tseng
          \inst{1}\inst{2}
          \and
          D.~Bockel\'ee-Morvan
          \inst{1}
	  \and
	  J.~Crovisier
          \inst{1}
	  \and
	  P.~Colom
          \inst{1}
	  \and
	  W.-H.~Ip
	  \inst{2}
          }

   \offprints{J. Crovisier\\
   \email{jacques.crovisier@obspm.fr}}

   \institute{Observatoire de Paris, 5 place Jules Janssen, F-92195 Meudon, France
   \and
   Institute of Astronomy, National Central University, 
   Chung Li 32054, Taiwan  
             }
	     
%
\date{Received 31 October, 2006; accepted 7 February 2007}


\abstract 
{} 
{We retrieve the H$_2$O expansion velocity in a number of comets,
using the 18-cm line shapes of the OH radical observed with the
Nan\c{c}ay radio telescope.} {The H$_2$O velocity is derived from the
large base of a trapezium fitted to the observed spectra.  This
method, which was previously applied to 9 comets, is now extended to
30 further comets.  This allows us to study the evolution of their
water molecule outflow velocity over a large range of heliocentric
distances and gas production rates.}
%
%
{Our analysis confirms and extends previous analyses.  The retrieved
expansion velocities increases with increasing gas production rates
and decreasing heliocentric distances.  Heuristic laws are proposed,
which could be used for the interpretation of observations of cometary
molecules and as a touchstone for hydrodynamical models.  The
expansion velocities retrieved from 18 cm line shapes are larger than
those obtained from millimetric observations of parent molecules with
smaller fields of view, which demonstrates the acceleration of the gas
with cometocentric distance.  Our results are in reasonable
quantitative agreement with current hydrodynamical models of cometary
atmospheres.}{}


\maketitle

\section{Introduction}
\label{sect:intro}

Due to the small gravity field of cometary nuclei, cometary
atmospheres are not captive and expand freely in the interplanetary
medium.  The expansion velocity of the atmosphere is a parameter of
crucial importance for the interpretation of cometary observations and
the modelling of cometary phenomena.  This velocity is not constant;
it is governed by the sublimation mechanism, photolytic heating of the
coma and collisions, so that case-by-case studies are in principle
necessary.  It is suspected to depend basically upon the heliocentric
distance, which governs photolytic heating, and the gas production
rate, which governs collisions (e.g., \citealt{bock-crov:1987,
comb+:2005}).  Heuristic laws have been proposed, but they need to be
checked and validated by hydrodynamical models and direct
measurements.

One of the best methods to determine this expansion velocity is the
observation of the shapes of molecular lines at radio wavelength,
taking benefit of the high spectral resolution of such observation.
These lines, when optically thin (which is practically the case for
all species except water), have purely Doppler profiles.  Therefore,
they directly trace the velocity distribution over the line of sight.
This has been applied to the millimetric and submillimetric lines of
molecules, such as HCN, CO, CH$_3$OH, H$_2$S, directly sublimated from
nucleus ices (e.g., \citealt{desp+:1986, schl+:1987, bive+:2002}).

One can also use the OH lines at 18 cm, for which an important
cometary database now exists.  However, OH is a secondary species,
coming from the photodissociation of water.  At photodissociation, the
OH radical is given an ejection velocity $V_d \approx 1$ km~s$^{-1}$
\citep{crov:1989}.  The kinematics and the space distribution of the
OH radical must be evaluated in the frame of a \textit{vectorial
model} \citep{comb-dels:1980, fest:1981}.

First attempts to deconvolve the OH line profiles in order to retrieve
the expansion velocity $V_p$ of the OH-parent used Monte Carlo
simulations \citep{bock-gera:1984, tacc+:1990}.  Another approach,
proposed by \citet[ hereafter Paper I]{bock+:1990}, derived the H$_2$O
velocity from the large base of a trapezium fitted to the observed
18-cm line shapes of the OH radical.  This method, which is much more
rapid than Monte Carlo simulations, was applied to 9 comets observed
with the Nan\c{c}ay radio telescope and checked over the Monte Carlo
method in specific cases.

In the present paper, we extend this work to 30 further comets
recently observed at Nan\c{c}ay, including C/1995 O1 (Hale-Bopp) which
was observed up to more than 4~AU from the Sun, and several comets
which were observed close ($\approx 0.5$~AU) to the Sun.  This allows
us to study the evolution of their water molecule outflow velocity
over a wide range of heliocentric distances and gas production rates.

Section \ref{sect:analysis} presents the database and the method of
analysis.  Section \ref{sect:results} presents the results and the
correlation between the gas expansion velocity, the heliocentric
distance and the gas production rate.  In Section
\ref{sect:discussion}, these results are discussed in the frame of
hydrodynamical models of cometary atmospheres and are compared with
other determinations of cometary expansion velocities.  Section
\ref{sect:conclusion} concludes.  Further details on this work are
given in \citet{tsen:2004}.

\begin{table}
\caption{Comets considered for the present analysis, listed by order
of perihelion dates.}
\centering
\begin{tabular}{lrc}
\hline\hline\noalign{\smallskip}
Comet                          & $N$ & ref. \\
\hline\noalign{\smallskip}
C/1982 M1 Austin               & $2$ & a) \\
C/1984 N1 Austin               & $2$ & a) \\
21P/Giacobini-Zinner (1985)    & $5$ & a) \\
C/1985 R1 Hartley-Good	       & $4$ & a) \\
C/1985 T1 Thiele               & $1$ & a) \\
1P/1982 U1 Halley              & $26$& a) \\
C/1986 V1 Sorrells             & $1$ & a) \\
C/1986 P1 Wilson               & $13$& a) \\
C/1987 P1 Bradfield            & $3$ & a) \\
C/1990 K1 Levy	               & $15$& b) \\
109P/1992 S2 Swift-Tuttle      & $10$& c) \\
C/1988 A1 Liller               & $1$ & d)  \\
23P/1989 N1 Brorsen-Metcalf    & $2$ & d)  \\ 
C/1989 Q1 Okazaki-Levy-Rudenko & $2$ & d)  \\ 
C/1989 W1 Aarseth-Brewington   & $1$ & d)  \\ 
C/1989 X1 Austin               & $3$ & d)  \\
C/1991 Y1 Zanotta-Brewington   & $1$ & d)  \\ 
C/1991 T2 Shoemaker-Levy       & $1$ & d)  \\
24P/Schaumasse (1993)          & $1$ & d)  \\ 
C/1993 Y1 McNaught-Russell     & $1$ & d)  \\ 
19P/Borrelly (1994, 2001)      & $8$ & d)  \\
45P/Honda-Mrkos-Pajdu\v{s}\'akov\'a (1996) & $1$ & d)  \\ 
C/1996 B2 Hyakutake            & $9$ & d)  \\
22P/Kopff (1996)               & $2$ & d)  \\
C/1996 Q1 Tabur                & $1$ & d)  \\
C/1995 O1 Hale-Bopp            & $28$& d)  \\
C/1998 J1 SOHO                 & $1$ & d)  \\
21P/Giacobini-Zinner (1998)    & $2$ & d)  \\
C/1999 H1 Lee                  & $7$ & d)  \\
C/1999 N2 Lynn                 & $1$ & d)  \\
C/1999 T1 McNaught-Hartley     & $3$ & d)  \\
C/2000 W1 Utsunomiya-Jones     & $1$ & d)  \\
C/2001 A2 LINEAR               & $9$ & d)  \\
C/2000 WM$_1$ LINEAR           & $3$ & d)  \\
153P/2002 C1 Ikeya-Zhang       & $8$ & d)  \\
C/2002 F1 Utsunomiya           & $2$ & d)  \\
C/2002 V1 NEAT                 & $6$ & d)  \\
C/2002 X5 Kudo-Fujikawa        & $3$ & d)  \\
C/2002 Y1 Juels-Holvorcem      & $3$ & d)  \\
\noalign{\smallskip}\hline
\end{tabular}
\smallskip
\begin{list}{}{}
\item $N$ is the number of samples for each comet. \\
\item a) \citet{bock+:1990}. \\
\item b) \citet{bock+:1992}. \\
\item c) \citet{bock+:1994}. \\
\item d) Present work.
\end{list}
\label{table:comet-list}
\end{table}

\begin{figure}
\centering
\includegraphics[width=\hsize,bb=54 309 558 760,clip]{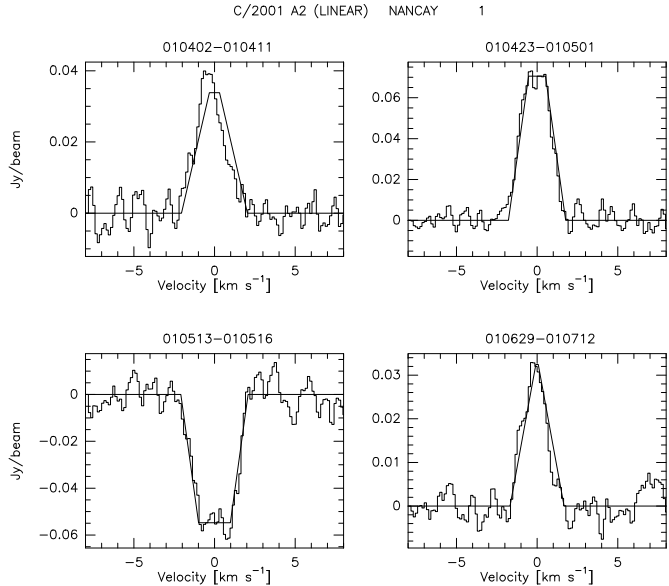}
\caption{A sample of OH spectra with their fitted trapezia: C/2001 A2
(LINEAR).}
\label{fig:trapeze}
\end{figure}

\begin{figure}
\centering
\includegraphics[height=\hsize,angle=270]{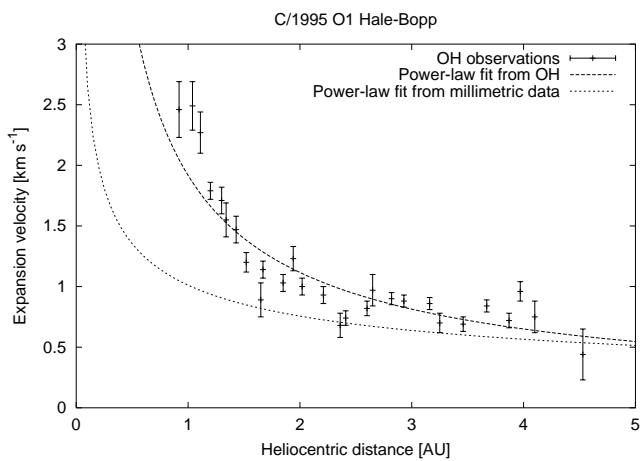}
\caption{Evolution of the expansion velocity $V_p$ as a function of
heliocentric distance $r_h$ for C/1995 O1 (Hale-Bopp).  Pre-perihelion
and post-perihelion data are combined.  The dashed line shows the
power-law fit (Eq.~\ref{eq1}) to the OH data.  The dotted line shows
the power-law fit (Eq.~\ref{eq2}) to the velocities derived from the
millimetric molecular lines \citep{bive+:2002}.}
\label{fig:HB}
\end{figure}

\begin{figure*}
\centering
\includegraphics[height=0.9\hsize,angle=270]{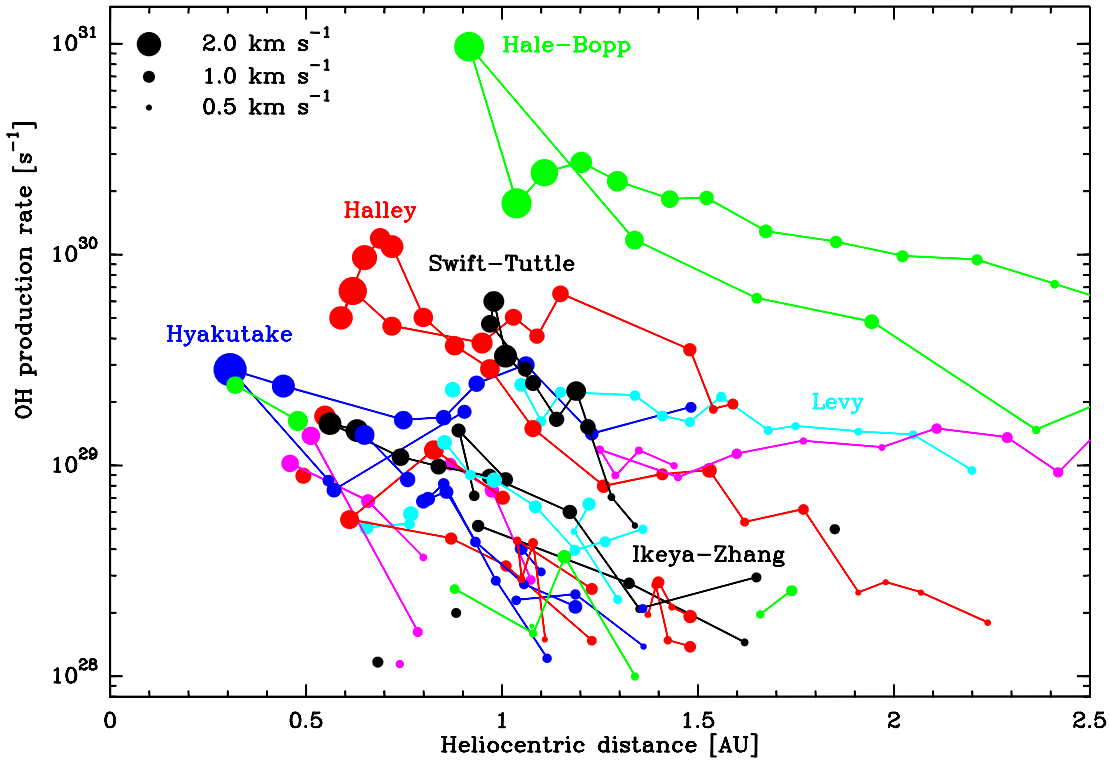}
\caption{Correlation between the retrieved water expansion velocity
$V_p$, heliocentric distance $r_h$, and OH production rate
$Q_\mathrm{OH}$.  The size of each circle is proportional to $V_p$. 
The well-studied comets 1P/Halley, C/1990 K1 (Levy),
109P/Swift-Tuttle, C/1996 B2 (Hyakutake), C/1995 O1 (Hale-Bopp) and
153P/2002 C1 (Ikeya-Zhang) are identified by their names.}
  \label{fig:correl}
\end{figure*}

\section{Analysis}
\label{sect:analysis}

The database of OH observations of comets at Nan\c{c}ay up to 1999 is
described in \citet{crov+:2002}\footnote{\sloppy See also\\ 
\texttt{http://www.lesia.obspm.fr/planeto/cometes/basecom/}}.  In
1995--2000, the Nan\c{c}ay radio telescope was upgraded
\citep{vdrie+:1997}, resulting in improved performances of the
instrumentation (the sensitivity was improved by a factor $\approx 2$
and a more versatile spectrometer was installed).  Cometary
observations after this upgrade are described by \citet[ and in
preparation]{crov+:2002-ACM}.  At 18~cm wavelength, the Nan\c{c}ay
radio telescope has an elliptical field of view of $3.5\arcmin \times
19\arcmin$, which corresponds to $r_1 \times r_2 = 76\,000 \times
415\,000$~km at $\Delta = 1$~AU and is equivalent to a circular field
of view of radius $r = \sqrt{r_1 r_2} = 178\,000$ km.

As explained in Paper I, a symmetric trapezium is fitted to the OH
line.  The large base of the trapezium is assumed to be $2(V_p +
V_d)$, where $V_p$ is the water expansion velocity and $V_d$ is the OH
ejection velocity upon water photodissociation.  As discussed in Paper
I, we assume $V_d = 0.9$ km~s$^{-1}$, which is close to the
theoretical value of 1.05 km~s$^{-1}$ \citep{crov:1989}.  Hence the
evaluation of $V_p$.  An example of OH spectra with their fitted
trapezia is shown in Fig.~\ref{fig:trapeze}.

As in Paper I, the OH production rates $Q_\mathrm{OH}$ are derived
using the OH inversion curve of \cite{desp+:1981} and the quenching
law and OH parameters of \cite{gera:1990}.  The model is fully
described in \citet[ Table 3, last column]{crov+:2002}.  The model
consistently uses the parent velocity derived from the trapezium fit
to compute the OH production rate; therefore, the production rates of
the present analysis may differ somewhat from those published by
\citet{crov+:2002} which were computed assuming $V_p = 0.8$
km~s$^{-1}$.

The list of comets investigated in the present study is given in
Table~\ref{table:comet-list}.  The analysis is performed on spectra
integrated over several days for which the signal-to-noise ratio is
sufficient (typically $> 10$).  The total number of samples is 190.  A
comprehensive tabulation of the data may be found in
\citet{tsen:2004}.

As an example, the expansion velocity as a function of heliocentric
distance $r_h$ is shown in Fig.~\ref{fig:HB} for C/1995 O1
(Hale-Bopp), for which the OH lines could be observed over a large
range of $r_h$.  The fitted power law is

\begin{equation}
    V_p = 1.917 (\pm0.055) r_h^{-0.78(\pm0.04)} \,\, \mathrm{km~s}^{-1}, 
\label{eq1}
\end{equation}

\noindent to be compared to 

\begin{equation}
    V_p = 1.125 (\pm0.015) r_h^{-0.42(\pm0.01)} \,\, \mathrm{km~s}^{-1}
\label{eq2}
\end{equation}

\noindent for the velocities derived from the millimetric molecular
lines \citep{bive+:2002} which were observed in a smaller field of
view (see discussion in Section \ref{sect:discussion}).  One must note
that in the range $0.92 < r_h < 4.57$ AU which is investigated here,
$Q_\mathrm{OH}$ varied by about three orders of magnitude, so that the
strong variation on $r_h$ implicitly includes a dependence on both
$r_h$ and $Q_\mathrm{OH}$.

\begin{figure}
\centering
\includegraphics[height=\hsize,angle=270]{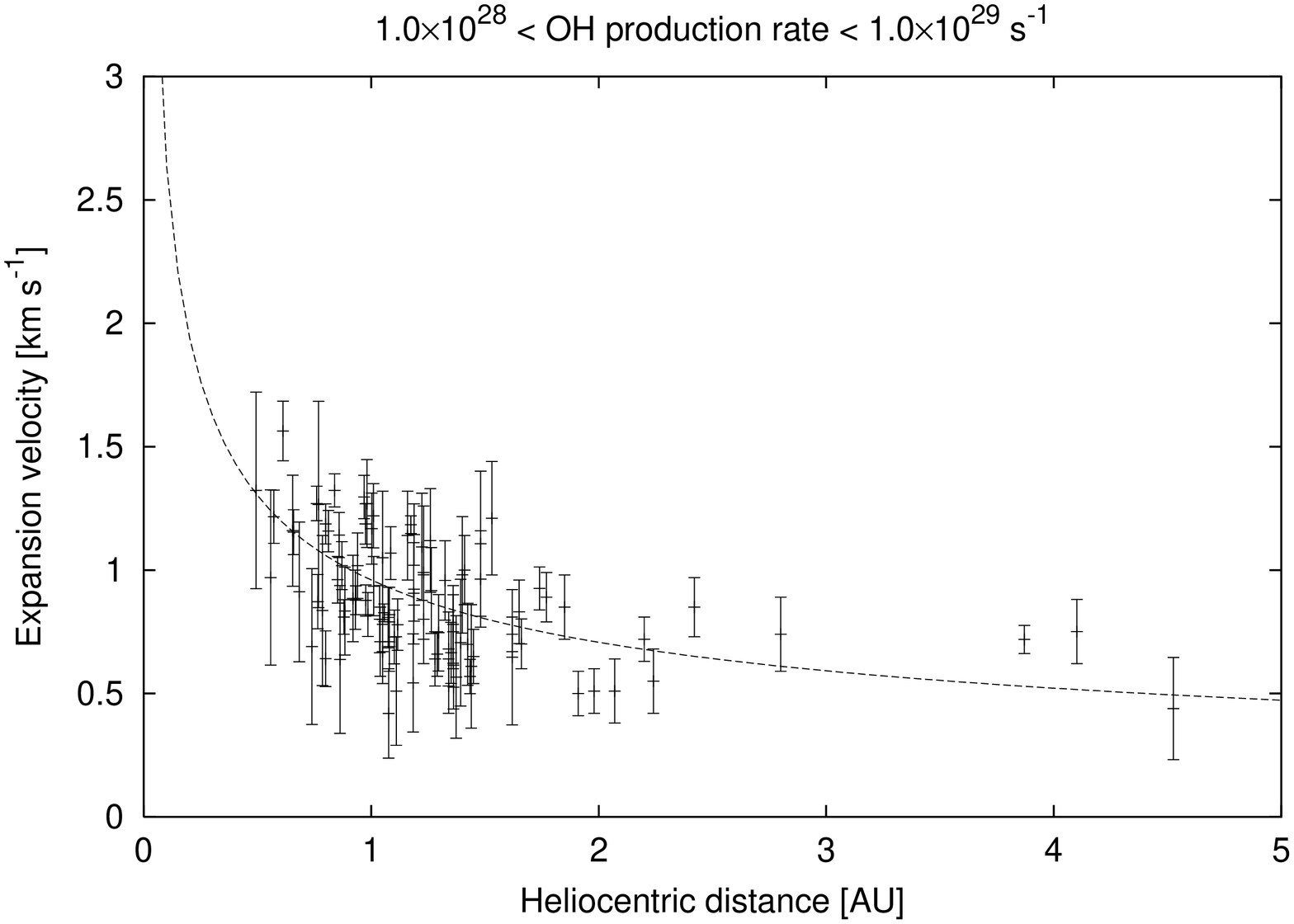}
\includegraphics[height=\hsize,angle=270]{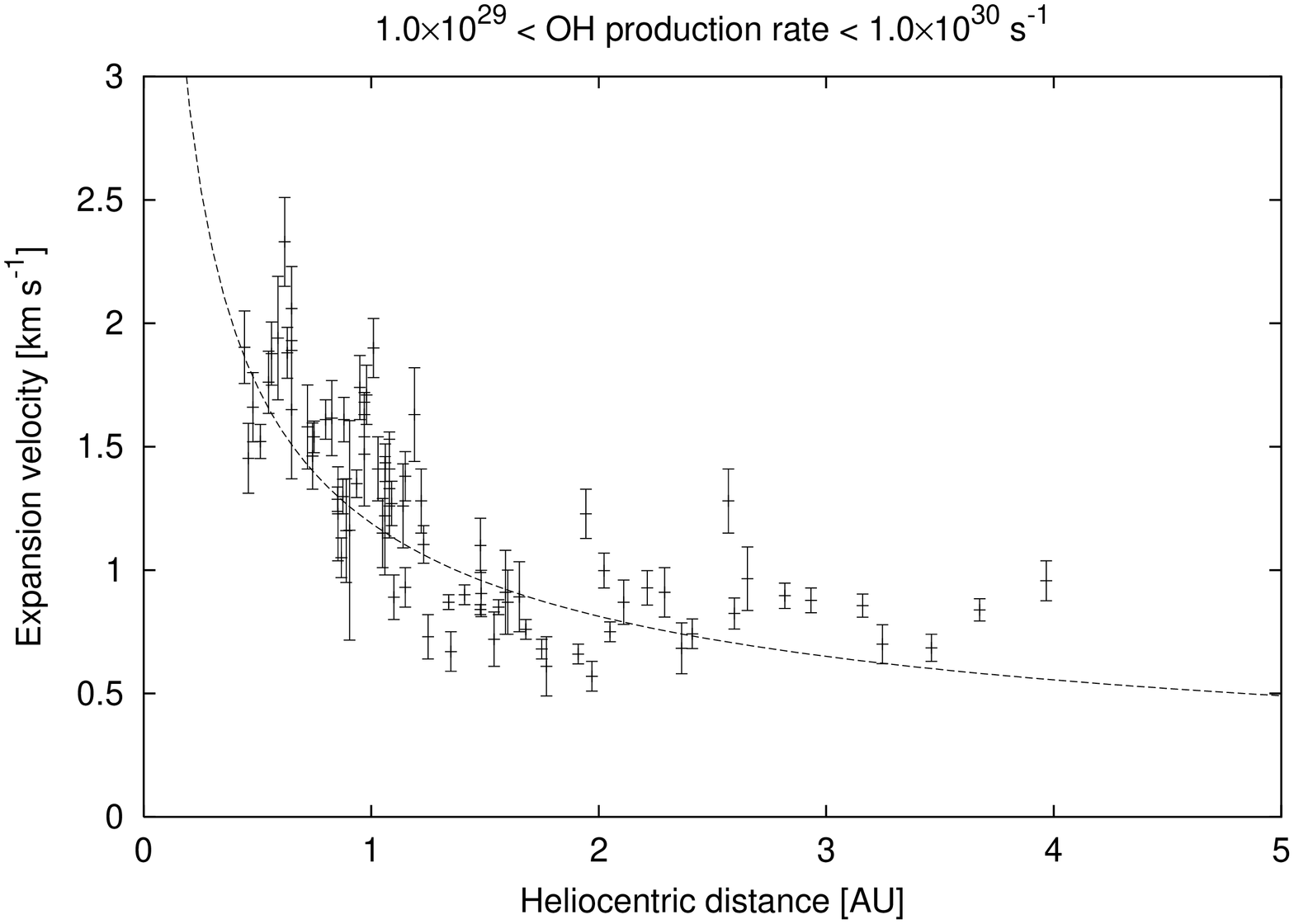}
  \caption{Evolution of the expansion velocity $V_p$ as a function of
  heliocentric distance $r_h$ for given $Q_\mathrm{OH}$ ranges.  The
  fitted power laws are listed in Table~\ref{table:power-laws}.}
\label{fig:evol1}
\end{figure}

\begin{figure}
\centering
\includegraphics[height=\hsize,angle=270]{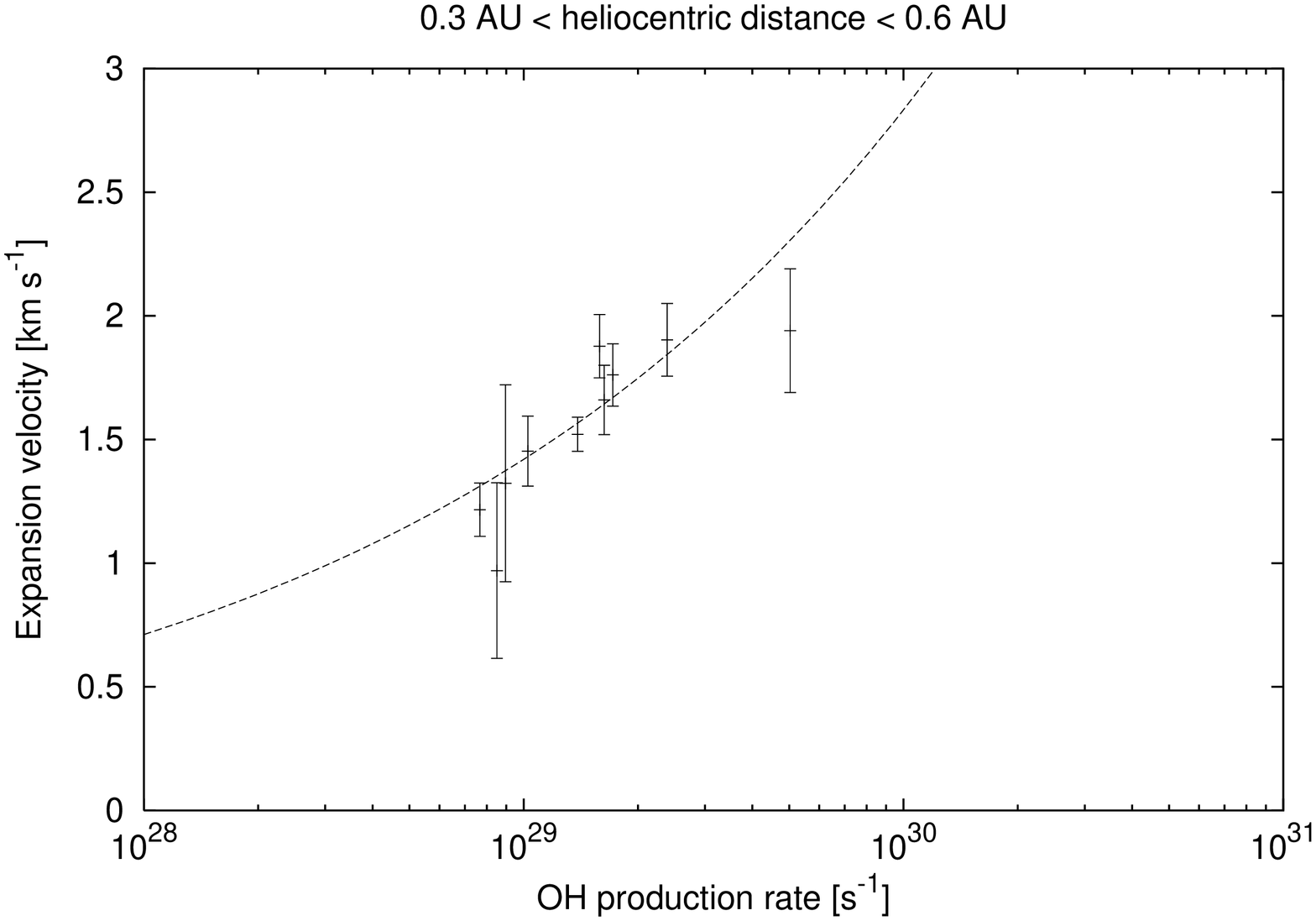}
\includegraphics[height=\hsize,angle=270]{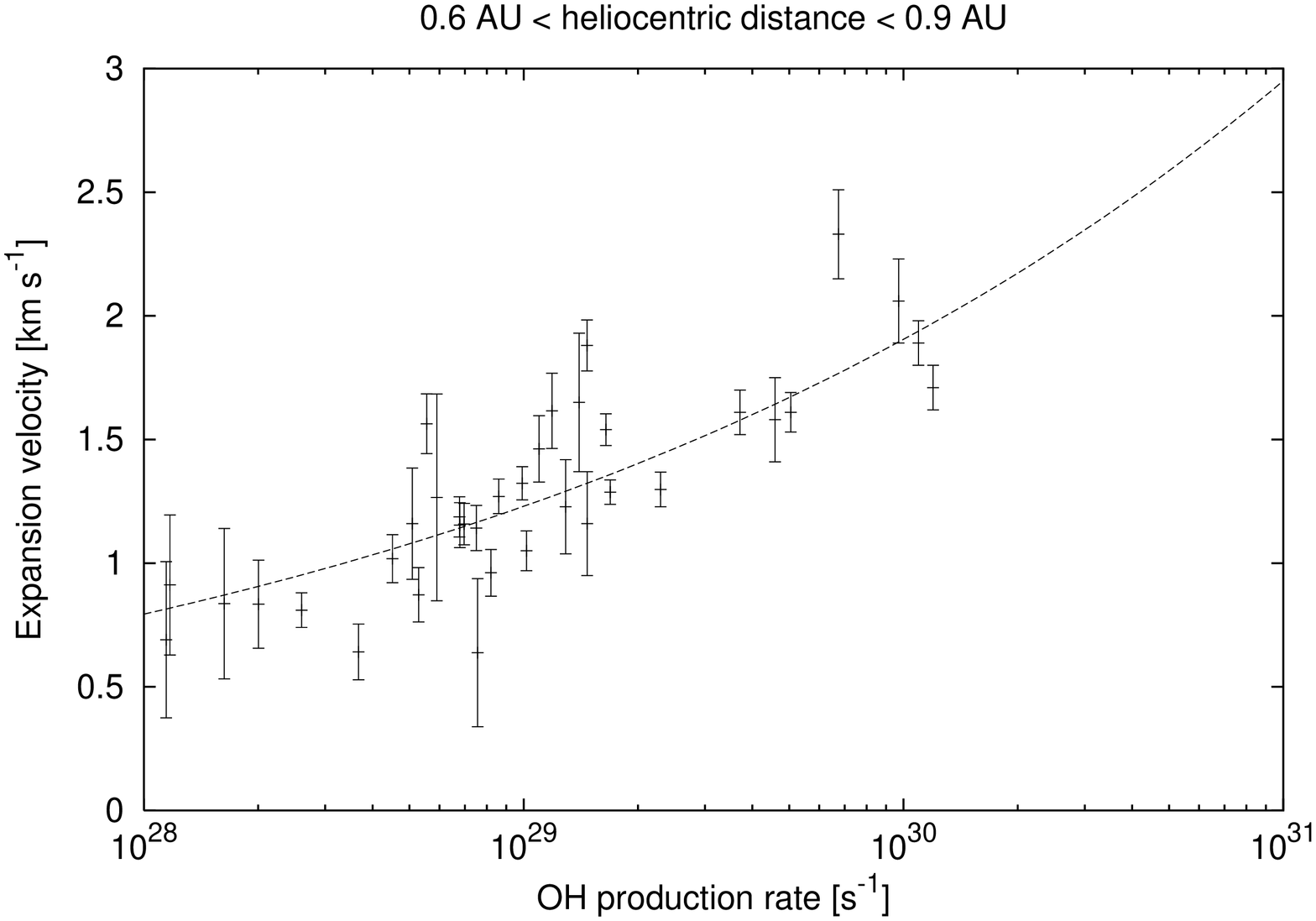}
\includegraphics[height=\hsize,angle=270]{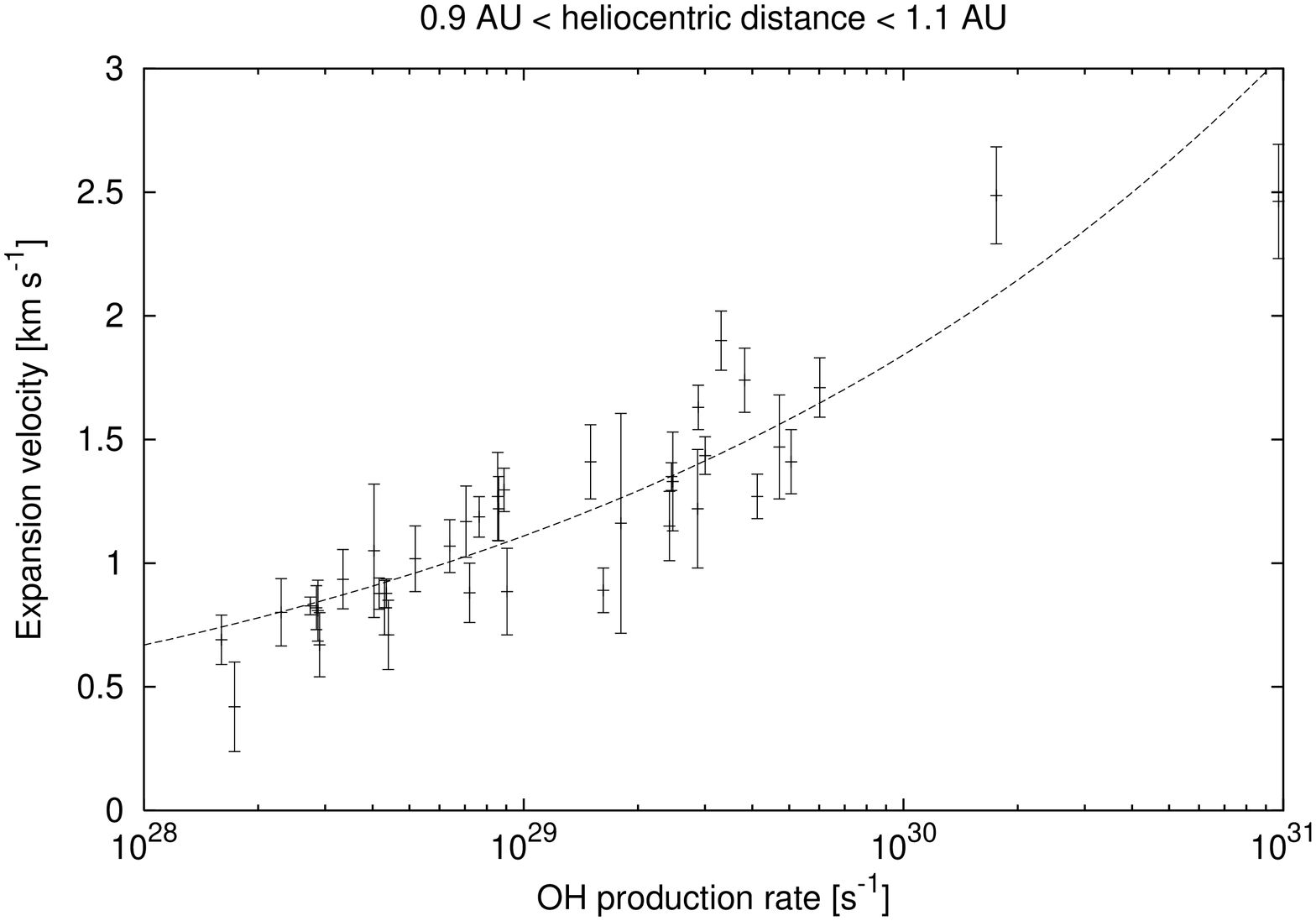}
  \caption{Evolution of the expansion velocity $V_p$ as a function of
  $Q_\mathrm{OH}$ for given heliocentric distance ($r_h$) ranges.  The
  fitted power laws are listed in Table~\ref{table:power-laws}.}
\label{fig:evol2}
\end{figure}

\addtocounter{figure}{-1}
\begin{figure}
\centering
\includegraphics[height=\hsize,angle=270]{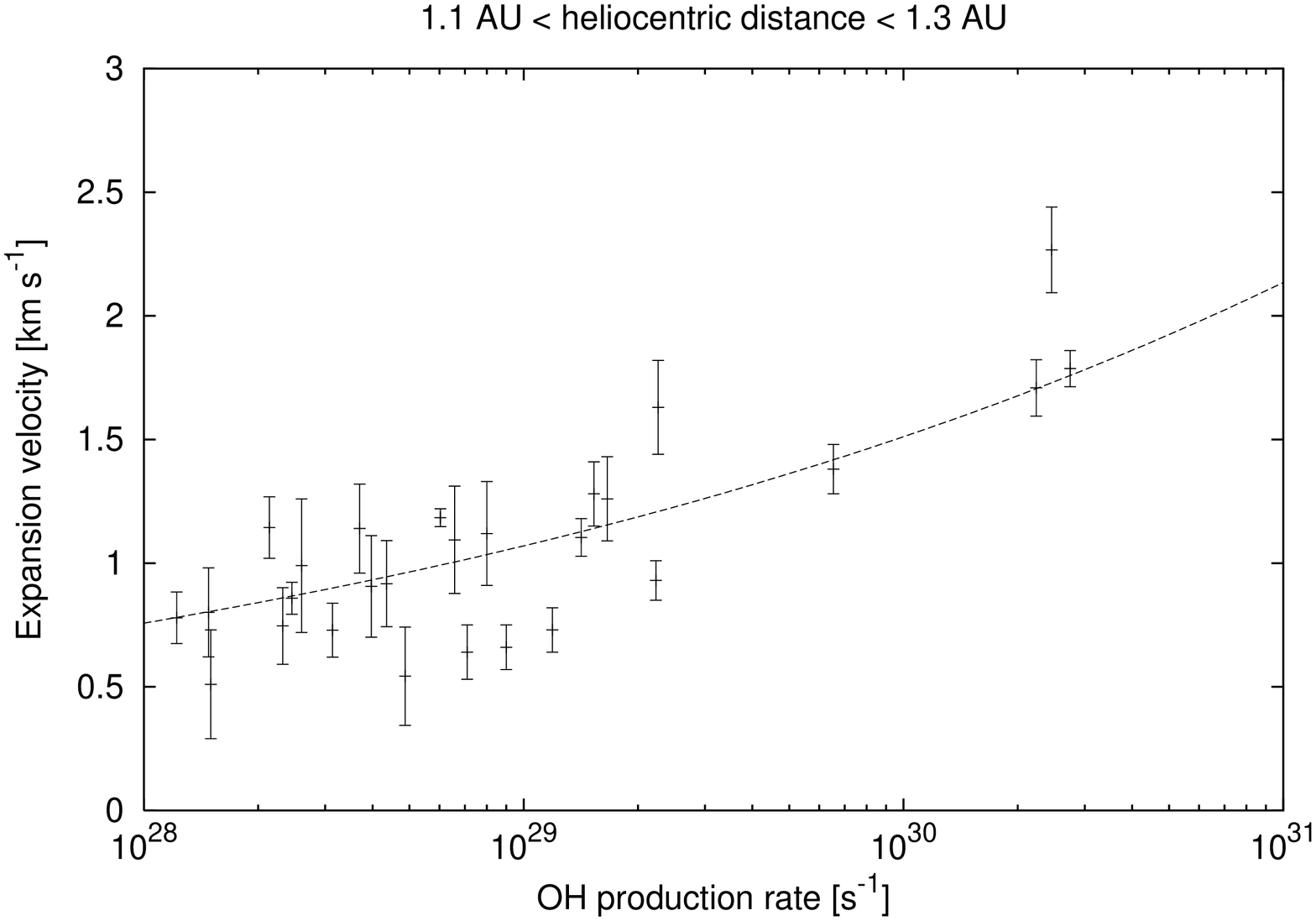}
\includegraphics[height=\hsize,angle=270]{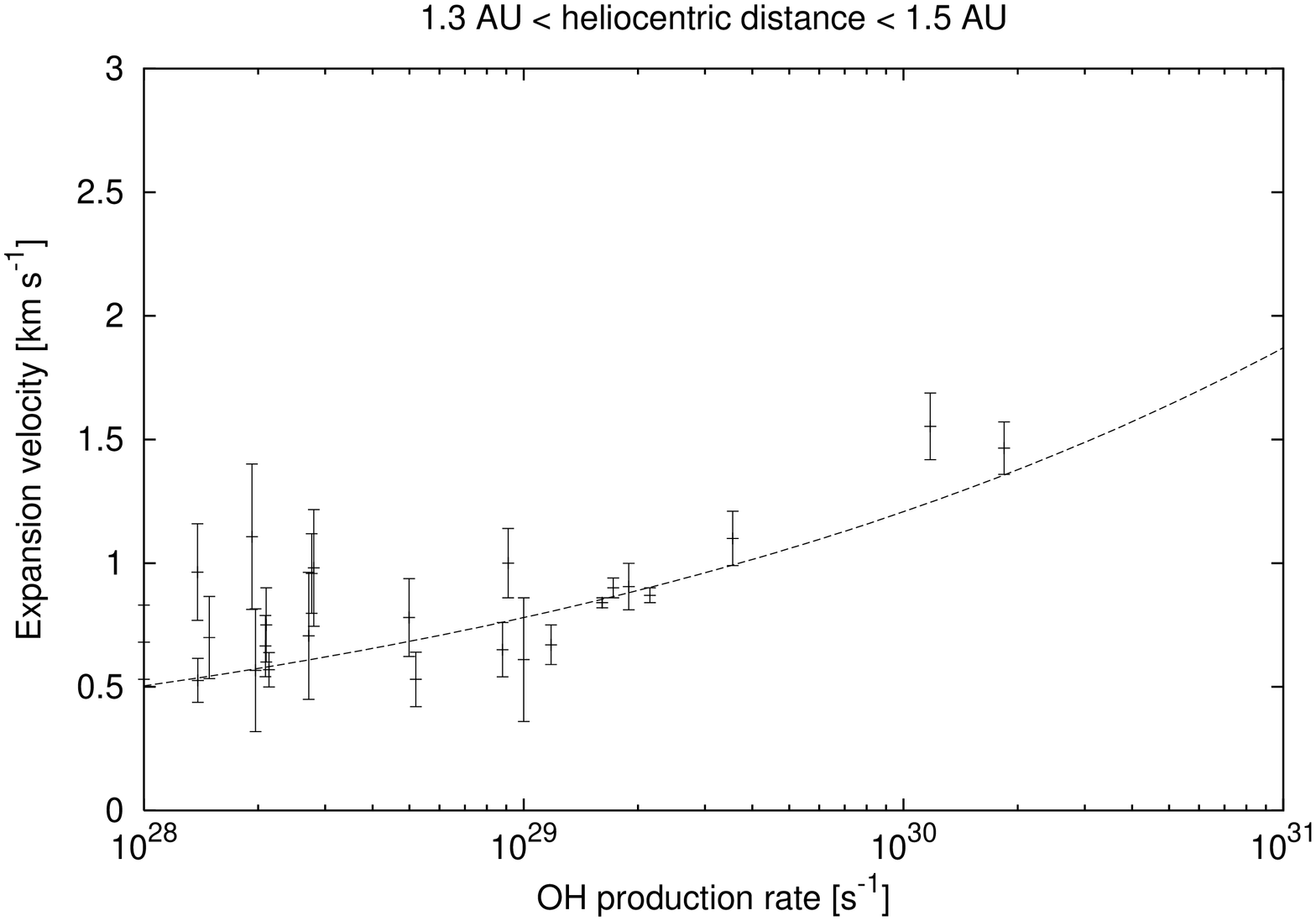}
  \caption{(continued)}
\label{fig:evol2bis}
\end{figure}

\begin{table}
\caption{Power-law fits to the H$_2$O expansion velocity $V_p$ (in
km~s$^{-1}$) obtained from the OH radio observations at Nan\c{c}ay for
different ranges of production rates $Q_\mathrm{OH}$ and heliocentric
distances $r_h$.  The corresponding data with plots of the fitted
power laws are shown in Figs \ref{fig:evol1}--\ref{fig:evol2bis}.}
\label{table:power-laws}
\centering
\begin{tabular}{cc}
\hline\hline\noalign{\smallskip}
$Q_\mathrm{OH}$ range & power-law fit \\
 {[s$^{-1}$]}         & [$r_h$ in units of AU]  \\
\noalign{\smallskip}\hline\noalign{\smallskip}
$10^{28}$ -- $10^{29}$ & $ {V_p} = 0.96\pm0.01\times{r_h}^{-0.44\pm0.04}$\\\noalign{\smallskip}
$10^{29}$ -- $10^{30}$ & $ {V_p} = 1.19\pm0.01\times{r_h}^{-0.55\pm0.02}$\\\noalign{\smallskip}
\hline
\hline\noalign{\smallskip}
$r_h$ range  & power-law fit \\
 {[AU]}      & [$Q_\mathrm{OH}$ in units of $10^{29}$ s$^{-1}$] \\
\noalign{\smallskip}\hline\noalign{\smallskip}
0.3 -- 0.6 & $ {V_p} = 1.42\pm0.05 \times {Q_\mathrm{OH}}^{0.30\pm0.06}$\\\noalign{\smallskip}
0.6 -- 0.9 & $ {V_p} = 1.23\pm0.02 \times {Q_\mathrm{OH}}^{0.19\pm0.01}$\\\noalign{\smallskip}
0.9 -- 1.1 & $ {V_p} = 1.11\pm0.02 \times {Q_\mathrm{OH}}^{0.22\pm0.01}$\\\noalign{\smallskip}
1.1 -- 1.3 & $ {V_p} = 1.07\pm0.02 \times {Q_\mathrm{OH}}^{0.15\pm0.01}$\\\noalign{\smallskip}
1.3 -- 1.5 & $ {V_p} = 0.78\pm0.02 \times {Q_\mathrm{OH}}^{0.19\pm0.02}$\\\noalign{\smallskip}
\hline
\end{tabular}
\end{table}

\section{Results}
\label{sect:results}

From the profiles of the 18-cm OH lines, we have investigated the
expansion velocity $V_p$ of cometary atmospheres for heliocentric
distances $r_h$ ranging from 0.3 to 4.6 AU and OH production rates
$Q_\mathrm{OH}$ from $10^{28}$ to $10^{31}$ s$^{-1}$.  Confirming the
results of Paper I, $V_p$ is found to consistently increase with
increasing $Q_\mathrm{OH}$ and with decreasing $r_h$, as expected
qualitatively from hydrodynamical models (\citealt{comb+:2005} and
references therein).  The observed $V_p$'s range from 0.5 to
2.5~km~s$^{-1}$.

An overview of the correlation between retrieved water expansion
velocities, heliocentric distances and OH production rates is shown in
Fig.~\ref{fig:correl} (an update of Fig.~14 in Paper I).  In order to
investigate separately the dependence of $V_p$ to $r_h$ and
$Q_\mathrm{OH}$, we have divided the data into various subsamples.
Power laws fitted to the different subsamples are listed in
Table~\ref{table:power-laws} and shown in Figs
\ref{fig:evol1}--\ref{fig:evol2}.

From these plots, we can see the existence of a threshold effect: the
expansion velocities are $\approx 0.8$~km~s$^{-1}$, insensitive to the
heliocentric distances from 1.5 to 5.0~AU for moderately active comets
with OH production rates from a few $10^{28}$ s$^{-1}$ to a few
$10^{29}$ s$^{-1}$.  The expansion velocities strongly depend on the
heliocentric distance for $r_h \le$ 1.5~AU. 

Errors reported in Table~\ref{table:power-laws} are the 1--$\sigma$
formal errors resulting from the fits to the data.  They do not include possible
systematic errors due to shortcomings in the modelling.  Given the limited
number of data, their arbitrary binning, and possible systematic effects, we do
not think that the differences of the exponent of the power law $r_h^\alpha$, in
the upper part of Table~\ref{table:power-laws}, are firmly established.  The
same for the exponent of $Q_\mathrm{OH}^\beta$ in the lower part of the table.
$\alpha = -0.5$ and $\beta = 0.2$ are probably the values to be retained.

Taking a step further, we tried to fit a combined power law
$r_h^\alpha Q_\mathrm{OH}^\beta$ to our data.  A unique law could not
be obtained for the whole domain covered by the observations.  The
results are shown in Table~\ref{table:comb-power-laws}.  For $r_h$ $>$
2 AU, $V_p$ barely depends upon $r_h$ and $Q_{\rm OH}$ as already
discussed.  The dependence with $Q_{\rm OH}$ is $\propto$ $Q_{\rm
OH}^{0.20}$ over the range 0.3--2 AU, as inferred from $Q_{\rm OH}$
power-law fits over restricted ranges of $r_h$
(Table~\ref{table:power-laws}).  Between 0.3 and 1 AU, the $r_h$
dependence is in $r_h^{-0.45}$, but the data suggest a steeper
heliocentric dependence in $r_h^{-1}$ between 1 and 2 AU. Power-law
fits in the subsamples [$10^{28} < Q_{\rm OH} < 10^{29}$~s$^{-1}$~; $1
< r_h < 1.5$~AU] and [$10^{29} < Q_{\rm OH} < 10^{31}$~s$^{-1}$~; $1 <
r_h < 2$~AU] are similar ($\sim$ $Q_{\rm OH}^{0.20} r_h^{-1}$), which
suggests that the difference in heliocentric evolution between $r_h$
$<$ 1 AU and $r_h$ $>$ 1 AU could be a characteristics of $V_p$
behaviour.  Note, however, that there is significant dispersion
between the data and the fitted curves (reduced $\chi^2$ between 2 and
3, see Table~\ref{table:comb-power-laws}).

\section{Discussion}
\label{sect:discussion}

The empirical laws derived from non-linear fits and listed in
Table~\ref{table:power-laws} can be used as a touchstone for
hydrodynamical models of cometary atmospheres.  They can also help us
to analyse cometary observations, and especially to derive more
reliable OH production rates from the 18-cm observations, when the
signal-to-noise ratio is not sufficient to apply the trapezium method
to the line shapes.

These results, however, should only be used for the range of
parameters ($r_h$, $Q_\mathrm{OH}$) pertaining to the data analysed in
the present work.  In particular, we caution the reader against
extrapolating the laws of Table~\ref{table:power-laws} to distant
comets.  At large $r_h$'s, cometary activity is governed by the
sublimation of CO and other hypervolatile species rather than that of
water, resulting in quite different temperatures and velocities at the
nucleus surface \citep{ip:1983}.

The top part of Table~\ref{table:power-laws} shows that, for a
restricted range of $Q_\mathrm{OH}$, the heliocentric dependence is
close to the law

\begin{equation}
V_p = 0.85 \times r_h^{-0.5} \,\, \mathrm{km~s}^{-1}
\end{equation}

\noindent which was proposed sometime (e.g., \citealt{coch-schl:1993,
budz+:1994}).  But the dependence upon $Q_\mathrm{OH}$ cannot be
ignored.  \citet{coch-schl:1993} proposed the law

\begin{equation}
V_p = 0.7 + 1.3 \times {Q_\mathrm{H_2O}}^{0.5}\,\, \mathrm{km~s}^{-1}
\end{equation}

\noindent for $r_h \approx 1$~AU from the results of Paper I, where
$Q_\mathrm{H_2O}$ is in units of $10^{30}$ s$^{-1}$.  This law appears
to be a poor fit to our data.

\begin{table*}
\caption{Combined $r_h$ and $Q_{\rm OH}$ power-law fits to the
expansion velocity $V_p$ (in km s$^{-1}$) for different ranges of
$Q_{\rm OH}$ and heliocentric distance.}
\label{table:comb-power-laws}
\centering
\begin{tabular}{cccc}
\hline\hline\noalign{\smallskip}
 $r_h$ range   & $Q_\mathrm{OH}$ range                & combined power-law fit                        & reduced $\chi^2$ \\
 {[AU]}        & [s$^{-1}$]                           & [$Q_{\rm OH}$ in units of 10$^{29}$ s$^{-1}$] & \\
\noalign{\smallskip}\hline\noalign{\smallskip}
 $0.3<r_h<1.0$ & $ 10^{28} < Q_\mathrm{OH} < 10^{30}$ & $V_p = 1.11\pm0.02 \times r_h^{-0.45\pm0.05} \times Q_\mathrm{OH}^{0.23\pm0.01}$ & 2.0 \\\noalign{\smallskip}
 $1.0<r_h<2.0$ & $ 10^{28} < Q_\mathrm{OH} < 10^{31}$ & $V_p = 1.17\pm0.02 \times r_h^{-0.99\pm0.05} \times Q_\mathrm{OH}^{0.18\pm0.01}$ & 3.3 \\\noalign{\smallskip}
 $2.0<r_h<4.6$ & $ 10^{28} < Q_\mathrm{OH} < 10^{30}$ & $V_p = 0.71\pm0.02 \times r_h^{-0.07\pm0.08} \times Q_\mathrm{OH}^{0.09\pm0.02}$ & 2.0 \\\noalign{\smallskip}
\hline
\end{tabular}
\end{table*}
   
Interestingly, photochemical heating is more sensitive to the
heliocentric distance than to the water production rate.  The heating
rate is directly proportional to $r_h^{-2}$ (through the water
photodissociation rate), but is weighted by the efficiency of the
thermalization process of fast hydrogens.  The size of the region
where the efficiency is significant scales proportionally to the water
production rate (e.g., \citealt{ip:1983, bock-crov:1987}).  Though
other processes, such as radiative cooling, do affect the
hydrodynamics of the coma, the steeper $r_h$ dependency (compared to
$Q_{\rm OH}$ variation) observed in our data is consistent with
photolytic heating being the main process controlling the gas velocity
in the outer coma.  The different heliocentric variations in the
0.3--1 AU and 1--2 AU domains remain to be explained.

We will now compare our results with other observations and with model
predictions.  For this prospect, we have to take into account the
effect of the field of view and of the geocentric distance $\Delta$.

The \textit{expansion velocity} of a cometary atmosphere is not a well
defined parameter.  In the inner, collisional coma where classical
hydrodynamics prevails, this velocity is progressively increasing with
distance $r$ to the nucleus, as a result of photolytic heating
\citep{comb+:2005}.  In the outer coma where the free-molecular flow
is governed by rarefied gas dynamics, the velocity distribution is not
maxwellian.

The observed line shapes are averages of all molecules present in the
instrumental field of view.  In the following, we will assume that the observed
$V_p$ is representative of molecules at a distance $r$ equal to the
field-of-view radius.  This assumption is justified by Monte Carlo simulations
of the HCN line shape (Paper I, Fig.~15).  Alternatively, the field of view
radius for 18-cm observations may be larger than the water scale length $l_p$
for photodissociation (typically $l_p \approx 10^5$ km at 1~AU).  In this case,
we may wish to adopt $r = l_p$.  This must be considered as an approximation.

We note that one would also expect that for very small fields of view,
the collisional region is sampled, in which OH is thermalized and
partakes the kinetics of water.  Then the trapezium model would fail.
An extreme observational case was that of comet C/1996 B2 (Hyakutake),
which was observed at $r_h = 1.06$ AU, $\Delta = 0.13$ AU, with
$Q_\mathrm{OH} \approx 3 \times 10^{29}$ s$^{-1}$.  We retrieved $V_p
= 1.4$ km~s$^{-1}$, right in the range of the laws of
Table~\ref{table:power-laws}.  Quite recently, we observed
73P/Schwassmann-Wachmann 3 at $\Delta = 0.08$ AU only, with $r_h =
1.0$ AU and $Q_\mathrm{OH} \approx 10^{28}$ s$^{-1}$; the retrieved
$V_p$ (0.8 km~s$^{-1Ñ}$) was also \textit{normal} (Crovisier et al.,
in preparation; these observations are not in the database analysed in
the present work).  This suggests that for both observations, the
field of view was still not small enough for a significant
thermalization of the OH radicals, or that this thermalization also
quenched the OH maser, so that the 18 cm observations are insensitive
to such OH radicals.

Complementary data are provided by millimetric and submillimetric
observations of line shapes of parent molecules.  $Vp$ is then close
to the line half-width at half maximum (\citealt{bive+:2002} adopt for
$V_p$ 90\% of the half-width, to account for thermal broadening).
Data for several molecules (especially HCN, CO, CH$_3$OH and H$_2$S)
are now available for many comets (e.g., observations at IRAM, SEST,
JCMT and CSO; \citealt{bive+:1999, bive+:2000,bive+:2002,
bive+:2006-AA}).  Direct observations of the water line at 557~GHz
from space with the \textit{SWAS} and \textit{Odin} satellites also
exist \citep{neuf+:2000, leca+:2003}, but their kinematic
interpretation is hampered by the saturation of this strong line.

The typical angular field-of-view diameter for millimetric
observations with the IRAM 30-m telescope is $17\arcsec$ at a
frequency of 145~GHz, corresponding to a linear radius $r = 6\,200$ km
at a geocentric distance $\Delta = 1$~AU. This is much smaller than
the elliptical field of view of the Nan\c{c}ay radio telescope
(equivalent to a radius of 178\,000 km).

\begin{figure}
\centering
\includegraphics[height=\hsize,angle=270]{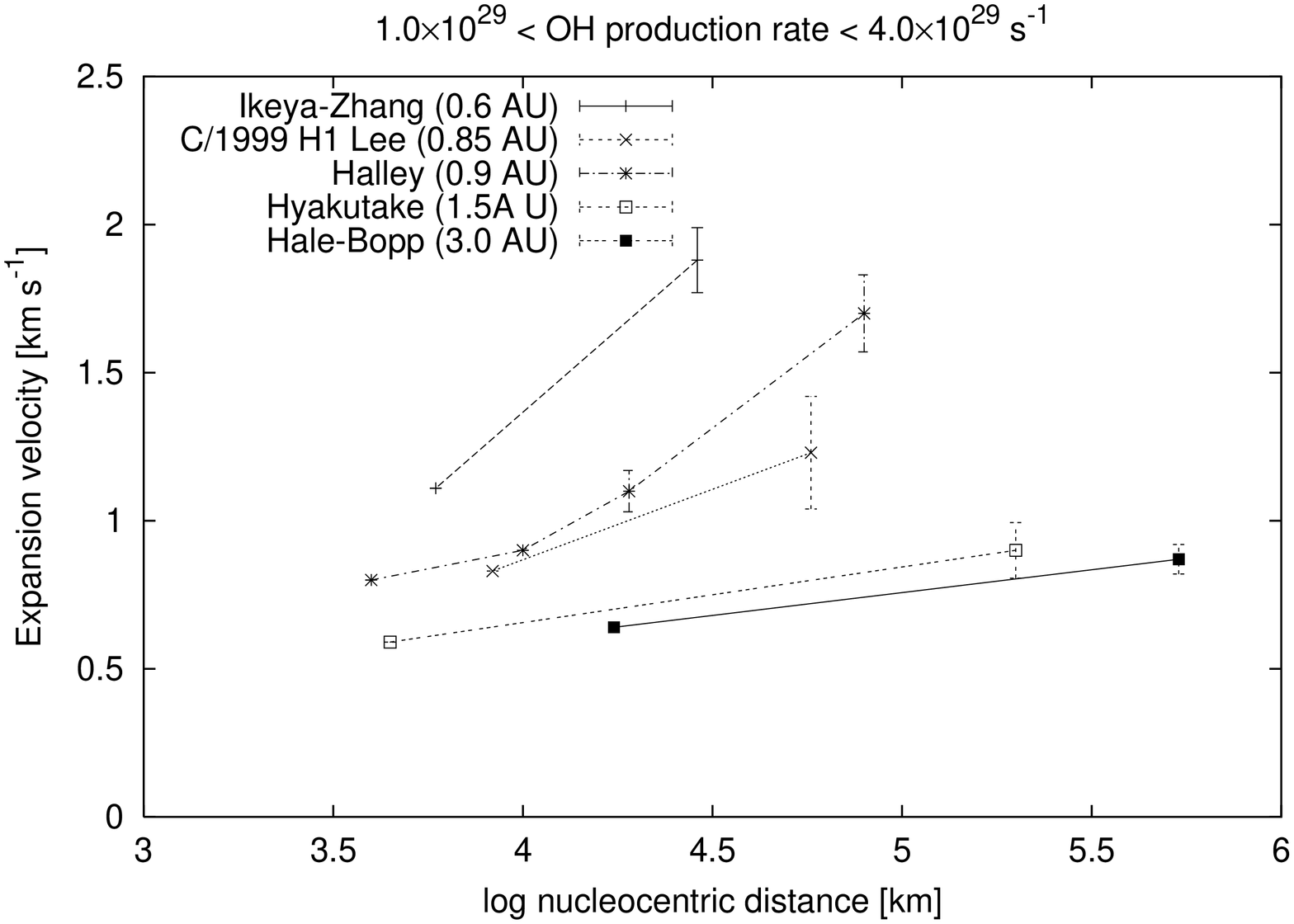}
  \caption{Field-of-view effects demonstrating the acceleration of the
  gas.  $V_p$ determined from the present work (right-hand points) or
  from millimetric observations (left-hand points) is plotted as a
  function of the cometocentric distance $r$ corresponding to the
  field of view.  See text for details.}
\label{fig:fov}
\end{figure}

As previously noted in, e.g., Paper~I, the $V_p$ determined from
parent molecules observed at millimetric wavelengths are
systematically smaller than those obtained from OH line shapes, with
larger fields of view.  This effect is shown in Fig.~\ref{fig:fov}.
Observations of comets 1P/Halley, C/1996 B2 (Hyakutake), C/1995 O1
(Hale-Bopp), C/1999 H1 (Lee) and 153P/2002 C1 (Ikeya-Zhang) were
selected at various $r_h$'s, chosen so that the gas production rates
were similar ($Q_\mathrm{OH}$ in the range 1--4 $\times 10^{29}$
s$^{-1}$).  Points on the right are $V_p$ determined from the present
work.  Points on the left are measurements from millimetric
observations \citep{bive+:1999, bive+:2000, bive+:2002,
bive+:2006-AA}.  For comet Halley, the various points from left to
right are from in situ measurements (\textit{Giotto},
\citealt{Lamm+:1987}), HCN (IRAM, \citealt{desp+:1986}), HCN (FCRAO,
\citealt{schl+:1987}) and OH (Nan\c{c}ay, present work).  As discussed
above, $r$ is evaluated as the field-of-view radius.  For all comets,
an increase of $V_p$ with $r$ is observed.  This increase is much more
important for comets close to the Sun (e.g., 153P/2002 C1
(Ikeya-Zhang) at $r_h = 0.6$ AU) than for distant comets (e.g., C/1995
O1 (Hale-Bopp) at 3 AU).  This clearly demonstrates the acceleration
of the gas with cometocentric distance, attributed to photolytic
heating.

Direct imaging of near-UV OH emission was made by \citet{harr+:2002}
in comet C/1995 O1 (Hale-Bopp).  From their analysis, $V_p = 2.3$
km~s$^{-1}$ was retrieved for $Q_\mathrm{OH} \approx 10^{31}$ s$^{-1}$
for cometocentric distances $\approx 10^6$ km.  This agrees well with
our measurement (2.4 km~s$^{-1}$).

\begin{figure}
\centering
\includegraphics[height=\hsize,angle=270]{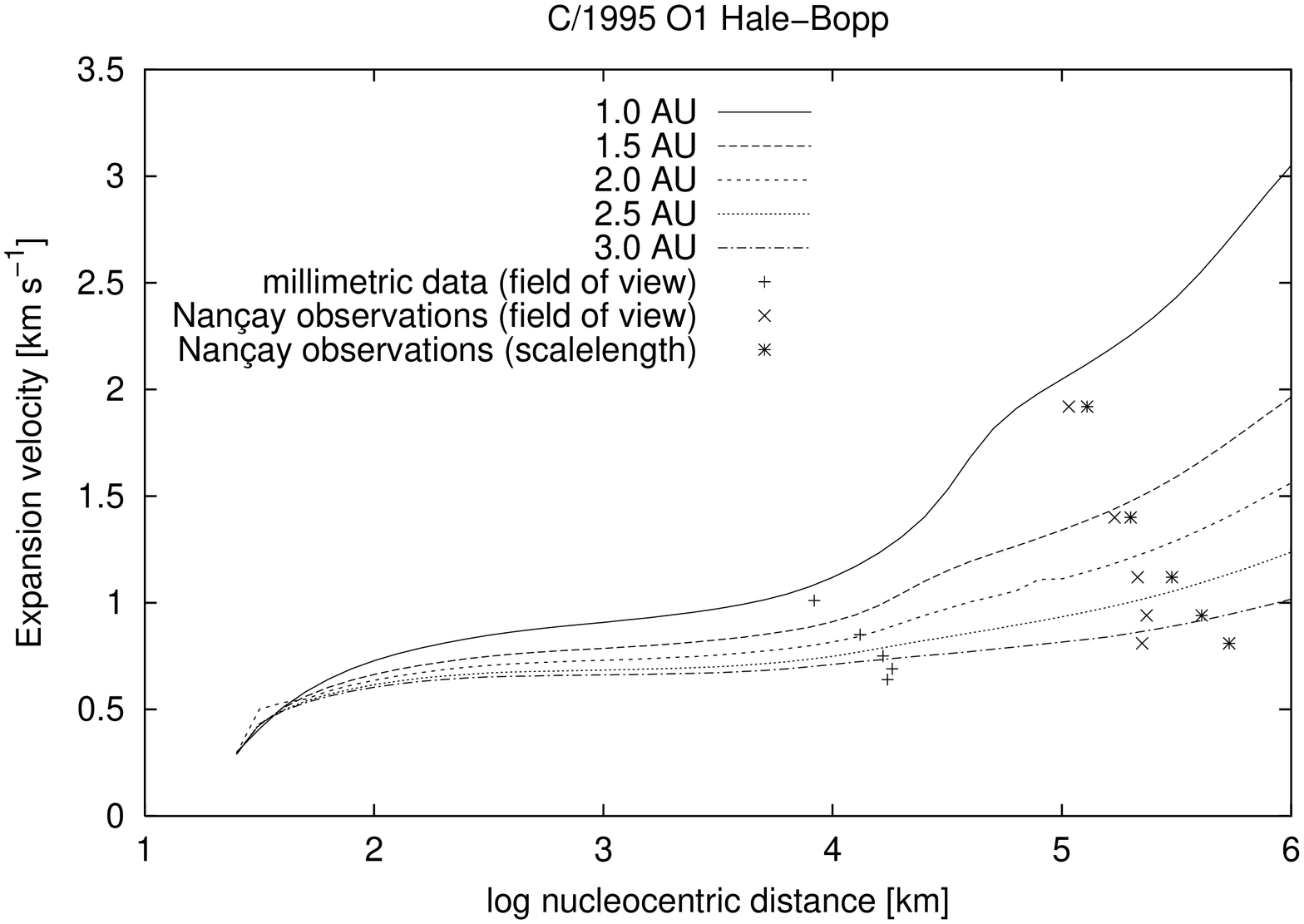}
  \caption{Model predictions of $V_p$ as a function of cometocentric
  distance, for comet C/1995 O1 (Hale-Bopp) pre-perihelion, adapted
  from Fig.~1 of \citet{comb+:1999}.  Observational determinations
  have been superimposed.  See text for details.}
\label{fig:model_combi}
\end{figure}

In order to achieve a quantitative comparison of these results with
models, we have plotted in Fig.~\ref{fig:model_combi} the predictions
of the hybrid kinetic/dusty gas hydrodynamical model of
\citet{comb+:1999} for comet C/1995 O1 (Hale-Bopp).  $V_p$ as a
function of cometocentric distance is plotted for five heliocentric
distances (1.0, 1.5, 2.0, 2.5 and 3.0 AU, all pre-perihelion) and the
corresponding observed water production rates (80, 20, 14, 4 and $4
\times 10^{29}$ s$^{-1}$, respectively, according to Combi et al.).
Observational determinations of $V_p$, from the present work (points
on the right) and from the millimetric observations of
\citet{bive+:2002} (points on the left), have been superimposed.  For
all these points, the cometocentric distances are again those
corresponding to the field-of-view radii.  We can see that the model
predictions lead to somewhat overestimated velocities.  Otherwise, the
variation of $V_p$ with $r$ and $r_h$ (and the related
$Q_\mathrm{OH}$) is remarkably well reproduced by the model.  The
discrepancy between model and observations may be attributed to:

\begin{itemize}
    
    \item flaws in the modelling of the heating/cooling processes (one
    can note that the model of \citet{comb+:1999} also predicts
    temperatures higher than those observed by \citealt{bive+:2002});
    
    \item the difficulty to model the transitional region between
    hydrodynamical flow and free-molecular flow (sophisticated Monte
    Carlo simulations are required; cf. \citealt{hodg:1990});
    
    \item an inaccurate estimation of the cometocentric distance to
    which the measured $V_p$ pertains (see discussion above);
    
    \item the presence of anisotropic outgassing, whereas models
    assume spherical symmetry.
    
\end{itemize}    

Finally, we note that, as already discussed in Paper I, there are some
observations, especially for distant, weak comets, leading to small
values of $V_p$ (0.5--0.6 km~s$^{-1}$) that can difficultly be
reconciled with hydrodynamical models.

\section{Conclusion}
\label{sect:conclusion}

We have extended the previous analysis of OH 18-cm line shapes from
the Nan\c{c}ay database (paper I) from 9 to 39 comets with the
following results:
    
\begin{itemize}
    
    \item This analysis confirms the increase of the coma expansion
    velocity $V_p$ with increasing gas production rate $Q_\mathrm{OH}$
    and decreasing heliocentric distance $r_h$.
    	
    \item The results are summarized in Tables \ref{table:power-laws}
    and \ref{table:comb-power-laws} which give power-law fits to the
    data.  These suggested laws can be used for testing hydrodynamical
    models of cometary atmospheres and for the interpretation of
    molecular observations, especially OH observations made with
    similar fields of views.
    
    \item The present results, compared to analyses of line shapes of
    parent molecules observed at millimetric wavelengths, yield
    significantly larger $V_p$'s.  This may be attributed to the much
    larger field of view of the 18-cm observations.  The two sets of
    data are complementary.  The increase of $V_p$ with the field of
    view demonstrates the acceleration of cometary gas with
    cometocentric distance.
    
    \item A threshold effect is observed: comets with small gas
    production rates (typically $Q_\mathrm{OH} <$ a few $10^{28}$
    s$^{-1}$) have similar expansion velocities (typically $V_p
    \approx 0.8$ km~s$^{-1}$).  Gas acceleration is inefficient for
    such weak comets.
    
    \item Our results are in reasonable agreement with current
    hydrodynamical models of cometary atmospheres.  Small
    discrepancies may be attributed to modelling issues such as
    inadequate treatment of the cooling/heating processes, or of the
    transitional region between hydrodynamical flow and free-molecular
    flow.
	
\end{itemize}

\begin{acknowledgements}

This work was done while W.-L. T. was guest of Observatoire de Paris.
Part of her work was supported by NSC grant 94-2111-M-008-033 and
Ministry of Education under the Aim for Top University Program NCU.
The Nan\c{c}ay Radio Observatory is the Unit\'e scientifique de
Nan\c{c}ay of the Observatoire de Paris, associated as Unit\'e de
service et de recherche (USR) No B704 to the French Centre national de
la recherche scientifique (CNRS).  Its upgrade was financed jointly by
the Conseil r\'egional of the R\'egion Centre in France, the CNRS and
the Observatoire de Paris.

\end{acknowledgements}

\bibliographystyle{aa}
\bibliography{ACM_AA_wendy}

\end{document}